\documentclass[times,11pt]{article}
\usepackage{graphicx}
\usepackage{hyperref}

\title{A Scalable Architecture for Harvest-Based Digital Libraries - The ODU/Southampton Experiments}
\author{Xiaoming Liu  \and Tim Brody \and Stevan Harnad \and Les Carr  \and Kurt Maly \and Mohammad Zubair \and Michael L. Nelson } 

\begin{document}

\maketitle
\begin{abstract}
This paper discusses the requirements of current and emerging applications based on the Open Archives Initiative (OAI) and emphasizes the need for a common infrastructure to support them. Inspired by HTTP proxy, cache, gateway and web service concepts, a design for a scalable and reliable infrastructure that aims at satisfying these requirements is presented. Moreover it is shown how various applications can exploit the services included in the proposed infrastructure. The paper concludes by discussing the current status of several prototype implementations.
\end{abstract}
\section{Introduction}
\protect\footnotetext[1]{X.Liu, K.Maly, M.Zubair are with Computer Science Department,Old Dominion University, USA}
\protect\footnotetext[2]{T.Brody, S.Harnad, L.Carr are with Intelligence, Agents, Multimedia Group, Electronics and Computer Science, University of Southampton, UK}
\protect\footnotetext[3]{M.L.Nelson is with NASA Langley Research Center, USA}
The Open Archives Initiative Protocol for Metadata Harvesting (OAI-PMH) is an important new infrastructure for supporting distributed networked information services. OAI promotes interoperability through the concept of metadata harvesting. The OAI framework supports Data Providers (DPs) and Service Providers (SPs). SPs develop value-added services based on the information collected from DPs. DPs are simply collections of harvestable metadata that may or may not contain additional services and content. This paper introduces the concepts of OAI-PMH proxies, OAI-PMH caches, OAI-PMH gateways as tools for the optimization of the functioning of the DP/SP model underlying the OAI-PMH.

The OAI-PMH uses HTTP-based request-response communication between a DP and an SP. The XML-formatted metadata is encoded in the HTTP response, which makes on-demand services possible. Using OAI-PMH, one DP may be harvested by any number of SPs, each possibly implementing different services. These service providers can interoperate using the multiple-resolution capability (one identifier is resolved to multiple instances) based on unique identifiers. In OAI-PMH, the metadata is distributed and replicated in many different places and potentially provides a highly redundant and fault-tolerant system.   

The Old Dominion University (ODU) Digital Library Group and the University of Southampton have been engaged in research focused on various OAI-PMH services. The goal of this research is to achieve interoperability, scalability and reliability of OAI-PMH services.

Over the past two years, ODU has developed the Arc \cite{arc}, Archon \cite{archon}, Kepler \cite{maly}, and DP9 \cite{dp9} services. Arc and Archon are federated searching services based on the OAI-PMH; they focus on the process of building a unified search interface over heterogeneous collections. Kepler is an end-user software that allows the individual to easily create and maintain a small, OAI-compliant archive. DP9 is a gateway service that allows general search engines (e.g. Google, Inktomi, etc.) to index OAI-PMH-compliant archives.

As part of the JISC/NSF Open Citation Project \cite{opcit}\cite{hitchcock1} (and previously the Open Journals Project), the University of Southampton Intelligence, Agents, Multimedia Group (IAM) has researched and developed methods for citation linking and analysis of the refereed scientific literature. Recent efforts have focused on CiteBase Search \cite{citebase}. CiteBase is an impact-ranking OAI-PMH SP, providing a federated searching service, citation linking (resolving author-provided references to their OAI-PMH-available targets), and citation analysis.
CogPrints \cite{cogprints} is an archive, hosted at Southampton, that allows authors to archive their articles (e-prints) for free in a freely accessible web service \cite{harnad1}\cite{cogprints}. The software behind CogPrints, ePrints.org \cite{eprintsorg}\cite{eprintsfaq}, has been developed by Southampton and released under Open Source. Now anyone, from individuals to institutions, may create their own self-archiving repositories.  ePrints.org is fully OAI-PMH-compliant.

In these efforts we notice challenges faced by OAI-PMH based applications, namely:

\begin{description}
\item[Data Provider and Metadata Quality]
During the testing of DPs, numerous problems were found. Not all archives strictly follow the OAI-PMH; many have XML syntax and encoding problems. The UPS Prototype \cite{ups} also found significant problems with metadata, however, with OAI-PMH, syntax for metadata is strictly defined (XML schema validation), problems still appear.  This suggests that some DPs do not strictly check their OAI-PMH implementations.   

\item[Server Availability]
The stability and service from DPs are difficult to predict since many factors may influence DP availability and efficiency \cite{nelson}.  If a large DP is periodically unavailable, this can be a serious problem for harvesting. A recent research points out that a significant number of DPs could not be harvested in the testing period  \cite{kent}.

\item[Scalability]
OAI-PMH harvesting is resource-expensive to DPs, partially because the HTTP responses are dynamically generated, and DPs may need to keep current harvest sessions (harvesting may take several days for a large data set). Besides steps taken by individual DPs to improve services, a general infrastructure is required.

\item[Linking Across Service Providers]
In OAI-PMH, several DPs may be harvested by many SPs, each providing different services for the same records. Cross-service linking and data sharing can be achieved by using the unique OAI identifiers. Unique identifiers also allow the detection of record duplication.

\end{description}

In this paper we discuss a joint effort between ODU and Southampton that addresses these problems using a variety of techniques. We present an architecture to achieve interoperability, scalability, and reliability by optimizing dataflow in DPs/SPs. This architecture introduces an OAI-PMH Proxy concept that could improve DP quality by fixing implementation problems just in time. An OAI-PMH Cache service improves data availability and avoids bottlenecks through Hierarchical Harvesting. An OAI-PMH Gateway translates operations from other resource discovery systems into operations in OAI-PMH and vice versa. We also discuss how to build a series of services such as cross-archive linking based on the suggested architecture. 

The remainder of this paper is organized as follows. In Section \ref{sect:related_work} we discuss related work. In Section \ref{sect:system_overview} we present an overview of the optimized model. We discuss each of the subsystems in Sections \ref{sect:data_provider}-\ref{sect:end_user_service}. In Section \ref{sect:working_applications} we discuss several working applications, and we summarize in Section \ref{sect:summary}.

\section{Related work} \label{sect:related_work}

\subsection{Global Research Archive}

The driving force behind the development of OAI-PMH was the need for a common method of federating heterogeneous  e-print archives into cross-archive search engines and other end-user services, e.g., the UPS prototype \cite{ups}.

A global research archive can start with e-print archives, perhaps subject-, institution-, or publisher- based. The software e-Prints.org has been developed to support author self-archiving \cite{harnad1}. This and other e-print archives can be harvested to form federated services, potentially a {\em{Global Research Archive}}, e.g., Scirus \cite{scirus.com}, Arc \cite{arc}, CiteBase \cite{citebase}, and My.OAI \cite{my.oai}.

This structured model for federated archives contrasts with CiteSeer \cite{giles98citeseer}, which is a web crawler that retrieves research articles from personal and institutional web sites, and automatically builds the bibliographic and reference data from the articles. There is no
interoperability model underlying CiteSeer.

\subsection{Caching and Replication}

HTTP proxy and cache distribute load, reduce network traffic and access latency, and protect the network from erroneous clients \cite{chankhunthod96hierarchical}. There are two basic approaches for web cache implementation: the passive cache only loads a data object as a result of a client's request to access that object, and an active cache employs some mechanism to prefetch data \cite{kroeger97exploring} in advance of a request by a client.

Mirror software is designed to duplicate a directory hierarchy between two machines. It avoids copying files unnecessarily by comparing the file timestamps and file sizes before transferring \cite{McLoughlin}. 

\subsection{Hierarchical Harvesting}

The earlier Harvest project explored the concept of the hierarchical arrangement of object caches and focused on the content extraction for general web documents \cite{bowman}. After OAI-PMH was released, both Arc and CiteBase explored the issues of hierarchical harvesting in OAI-PMH SPs. The Open Digital Library project \cite{suleman} and the OAI ``result set filtering'' whitepaper \cite{Vesely} explored using the OAI-PMH sets concept for OAI-PMH metadata filtering.

\subsection{Unique Identifiers}

Identifiers are a powerful tool for communication within and between communities. For example, the Handle system \cite{sun} and DOI (Digital Object Identifier) \cite{paskin} provide a mechanism for implementing naming systems for arbitrary digital objects. The multiple-resolution capability becomes important in OAI-PMH community, as metadata may be widely replicated and modified, and many different services will be implemented on the same metadata records. An ``intelligent'' resolution service should be able to deliver different outcomes to a resolution request dependent on user-specified requirements \cite{herbert}\cite{herbert1}. 

In OAI-PMH,  unique identifier unambiguously identifies an item within a repository, the format of the unique identifier must correspond to that of the URI syntax, Individual communities may develop community-specific URI schemes for coordinated use across repositories.

However, the unique identifiers may conform to a recognized URI scheme with greater scope. Especially the \textit{oai-identifier} schema describes a specific, recommended implementation of unique identifiers which repositories may adhere to; \textit{oai-identifiers} should have global scope and guaranteed global uniqueness. The \textit{oai-identifier} has been widely accepted in implementation of OAI-PMH 1.1 and is further refined in the version 2.0 of the protocol by introducing a globally unique OAI URN.  Both the ODU and Southampton implementations use the \textit{oai-identifier} schema and rely on its uniqueness.

\subsection{Citation Linking}

Citation linking is the general term for hypertext linking the reference lists (the bibliography) in research articles to the cited articles. In recent years, citation linking has been extensively developed \cite{hitchcock1} \cite{caplan} \cite{bergmark1} \cite{herbert}. With the wide acceptance of OAI-PMH, new challenges are raised about cross-archive (i.e. cross collection) linking and cross-service linking. With various DPs providing metadata of different qualities and formats, cross-archive linking is necessary to integrate them into one unique linking environment. Similarly, the distributed and highly redundant OAI-PMH architecture allows different services to be built which, with context sensitive and dynamic cross-service linking \cite{herbert} \cite{herbert1}, could potentially be integrated.

Such integrated services might provide citation analysis for forward links (to articles that have referenced the current article), impact factors, co-citation analysis, and novel navigation methods \cite{carrchen}.
\subsection{Distributed Search vs. Harvest}
There are two ways to implement DL interoperability, through distributed search or harvest. The distributed searching approach is studied in NCSTRL, Stanford Infobus \cite{Baldonado}\cite{davis}\cite{bowman}. Currently, the harvesting approach has better scalability and is a realistic approach for a large number of existing DLs. It is studied in Harvest [6] and is the basis of OAI-PMH.

\section{System Overview} \label{sect:system_overview}

\begin{figure}[ht]
\centering
\includegraphics[]{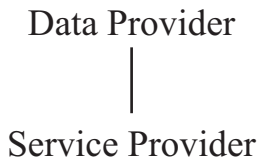}
\caption{Basic OAI-PMH Harvesting Model}
\label{fig:harvest_model}
\end{figure}

The need for an optimized model is motivated by the major challenges faced in a basic OAI-PMH model, which can be seen in Figure \ref{fig:harvest_model}.  The basic structure of OAI-PMH supports two roles: the SP and the DP.  Multiple SPs may harvest multiple DPs at the same time.  If one DP has implementation problems (e.g., XML encoding), all SPs have to address these problems.  If one DP is unavailable, all SPs have to wait until the DP comes up again, even if some SPs have already cached the data from the DP. 

\begin{figure}[ht]
\centering
\includegraphics[]{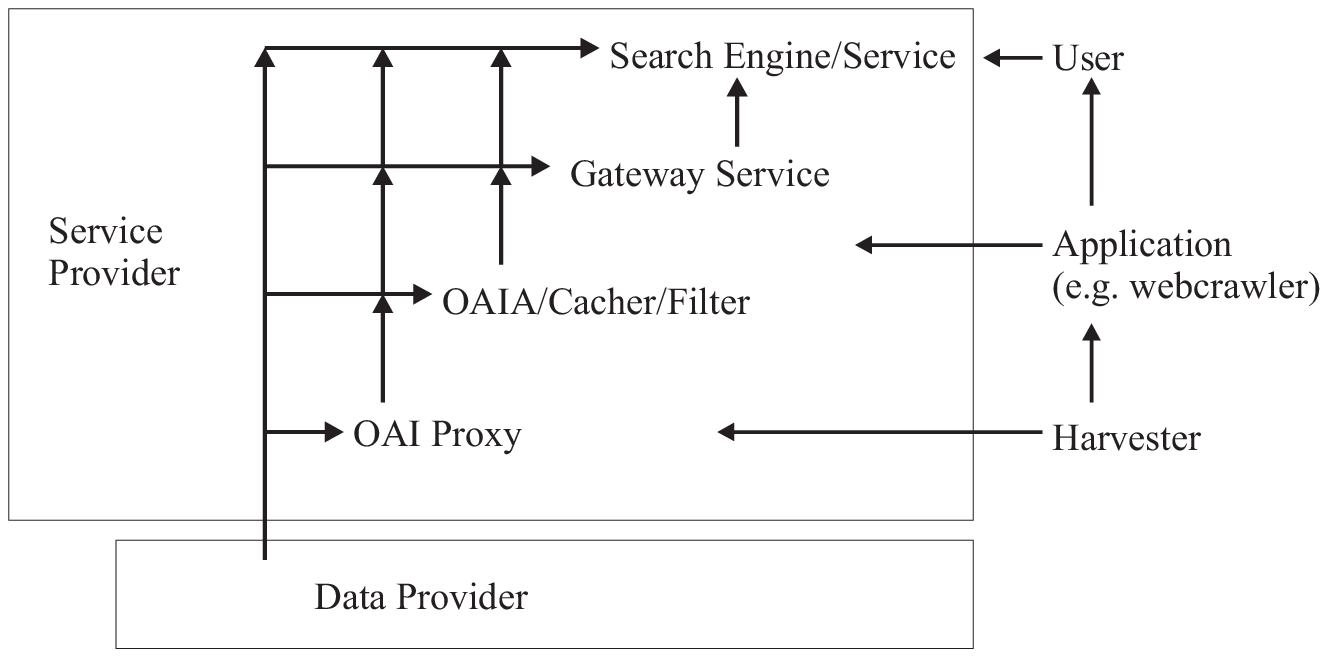}
\caption{Hierarchical Harvesting Model}
\label{fig:multilevel_model}
\end{figure}

Figure \ref{fig:multilevel_model} illustrates the optimized model based on the hierarchical harvesting. An OAI-PMH proxy dynamically forwards requests DPs with value-added services. For example, it can dynamically fix common XML encoding errors and translate between different OAI-PMH versions. An OAI-PMH cache caches metadata and can filter and refine them before exposing them to SPs. It also serves as a simple cache that reduces the load on source DPs and improves DP availability. An OAI-PMH gateway can convert the OAI-PMH to other protocols and applications. For example, the gateway could provide value-added services like automatic citation extraction, or conversion between different protocols (SOAP, Z39.50) and OAI-PMH. An end-user service will present various services such as search and citation linking. Figure \ref{fig:multilevel_model} illustrates how each layer may fetch data from any of its lower layers depending on availability and service type.

\section{Data Provider} \label{sect:data_provider}

The OAI-PMH is designed as a low-barrier and structured method for DPs to expose hidden content  to add-on services. (We define ``hidden'' as both data that cannot be accessed via the Web, and data that can only be accessed through proprietary web interfaces)

A DP, using OAI-PMH, exposes metadata for the resources for which it is responsible. Any (or many) metadata formats may be exposed for any given resource, although to be OAI-PMH compliant the DP must support Dublin Core.

A DP does not necessarily provide any added value from the end-user's perspective; it simply gives OAI-PMH compliant harvesters access to the raw metadata in the repository. However, a richer metadata resource allows more interesting services to be built on top of it. For example, DPs that are collections of research literature (articles with bibliographies), may export the resource's reference list as well as the current authors, title, and so on.

The first step to building value-added services on repositories is harvesting their metadata. Where there are OAI-PMH implementation faults, SPs must try to work around these problems, or, as we suggest, use a separate layer to fix errors with the DP's implementation: an OAI-PMH Proxy.

\section{OAI-PMH Proxy} \label{sect:oai_proxy}

From a harvester's point of view, the most convenient solution to incorrectly implemented DPs is to place a layer (i.e. a proxy) over source repositories that can be trusted to provide correct responses to the harvester's requests. The proxy can protect the network from erroneous and malicious clients, for example, a proxy can serve as the single access point for outside world to DPs inside a firewall.

An OAI-PMH proxy can either act as an HTTP proxy or be OAI-PMH-specific. As an HTTP proxy, it effectively becomes a transparent layer accepting HTTP requests and responding with HTTP responses. As an OAI-PMH-specific proxy, it must re-write request URLs; for an example of mapping a given subdirectory to a source baseURL, see Figure \ref{tab:oai_specific}.

\begin{figure}[ht]
\centering
\begin{tabular}{|l|l|}
	\hline
Request URL & Wrapped URL \\
	\hline \hline
oai-proxy/cgi/proxy/cogprints & cogprints.soton.ac.uk/perl/oai \\
	\hline
oai-proxy/cgi/proxy/bmc & www.biomedcentral.com/oai/1.1/bmcoai.asp \\
	\hline
\end{tabular}
\caption{OAI-PMH-Specific Style Proxy Requests}
\label{tab:oai_specific}
\end{figure}

An OAI-PMH proxy will fix the following errors:

\begin{description}
\item[Character Encoding] OAI-PMH uses the Unicode's UTF-8 character encoding to support international character sets by using multiple bytes for non-English characters. As an OAI-PMH response is received from a repository, the proxy can replace any faulty character encoding that would normally cause an XML parser to fail.
\item[XML Encoding] The mark-up characters used in XML must be encoded when used in string data. Similar to recent web browsers, the proxy can use heuristics to determine whether a mark-up character is actually part of mark-up, or should be encoded.
\item[XML Mark-Up] XML requires that all arguments to XML elements are quoted, and heuristics can be used to fix some quoting errors. An advanced proxy may be able to fix missing, or out-of-order, XML elements; however, this can easily lead to untraceable errors.
\item[Protocol Errors] A proxy can check the validity of the response it receives. It can check that the schemas used in the response are the ones used by the OAI-PMH, and that the XML can be validated against those schemas.
\end{description}

Whereas character and XML encoding problems can be fixed in a stream, more complex errors require caching the entire OAI-PMH response. This requirement leads naturally to an OAI-PMH Cache that is capable of storing and fixing OAI-PMH responses.

\section{OAI-PMH Aggregation and Caching} \label{sect:oai_aggregator}

OAI-PMH caching/aggregation (an OAIA) expands the DP/SP model with a middle layer that aggregates/caches responses from source DPs. This feature provides benefits to both aggregated DPs, and SPs that choose to harvest from the OAIA.

\subsection{Caching Data Providers}

A DP can be harvested by any number of services. Frequent harvesting by many services can result in a significant overhead for providers, who can range from small collections running on simple web-scripts to large multi-million record collections running on powerful databases. A small provider may have inefficient systems or few resources, and a large provider may not appreciate an unreliable harvester continually requesting very large data sets.

An OAIA solves this problem by mirroring DP's metadata, so a few large aggregators (efficiently implemented) can serve many other OAI-PMH services, including downstream aggregators.

\subsection{Aggregation}

When deciding which repositories to harvest from, an SP must consider whether the repository content is relevant, whether the repository is reliable, and how often the SP should check that repository for updates.

A hierarchical OAIA structure can reduce this problem by avoiding duplication of the efforts of many SPs. For example, an SP may want to provide an index of all the music manuscript repositories of a given country. That SP can then expose the aggregated collection to an international SP, saving the international SP the effort of harvesting from every repository in every country.

\subsection{Advantages Over HTTP Caching}

An OAIA is similar to an HTTP cache; specifically, they both distribute load away from the server (the DP) and closer to the client (the SP).

An OAIA is, however, an active cache - it requests new records from the known repositories in advance. This means a repository's records will always be available from the cache to downstream harvesters, even if the repository itself is unavailable.

An important role for an aggregator is providing quick access to many, smaller collections. By prefetching records from its source repositories, an OAIA can provide a downstream harvester with all the aggregated records in one session.

\subsection{Datestamping}

Incremental harvesting in OAI-PMH uses datestamping; that is, a harvester only needs to request records that are new or have changed since the last time it checked the repository.

With hierarchical harvesting the OAIA must update the datestamp when it harvests a record - because the record is ``new'' to the OAIA. When a downstream harvester harvests from the OAIA, it will receive all the new records in the OAIA, even if the original datestamp of the record was before the date of harvest.

\subsection{Identifiers}

An OAIA can either maintain or change the \textit{oai-identifiers} for records that it harvests (and reexports).

By maintaining the record's  \textit{oai-identifier}, the OAIA can become a nearly transparent layer in a hierarchical system (nearly transparent, because it introduces a delay between a record being created by DPs, and it being harvested from the OAIA). Maintaining \textit{oai-identifiers} allows a harvester to change sources without causing inconsistent records.

OAI-PMH 2.0 introduces a provenance schema for use in the optional ``about'' field of records (``about fields'' are for things that describe the metadata record). This schema allows the history of the record to be recorded: it stores the details of each OAI-PMH service that the record passes through as it goes down the hierarchy, from repository to eventual end-user service.

Provenance can be used to check the originality of the record, identify the change history of the record when it travels around the system, and to extract the original datestamp.

\subsection{Identifier Collisions}

\begin{figure}[ht]
\centering
\includegraphics[]{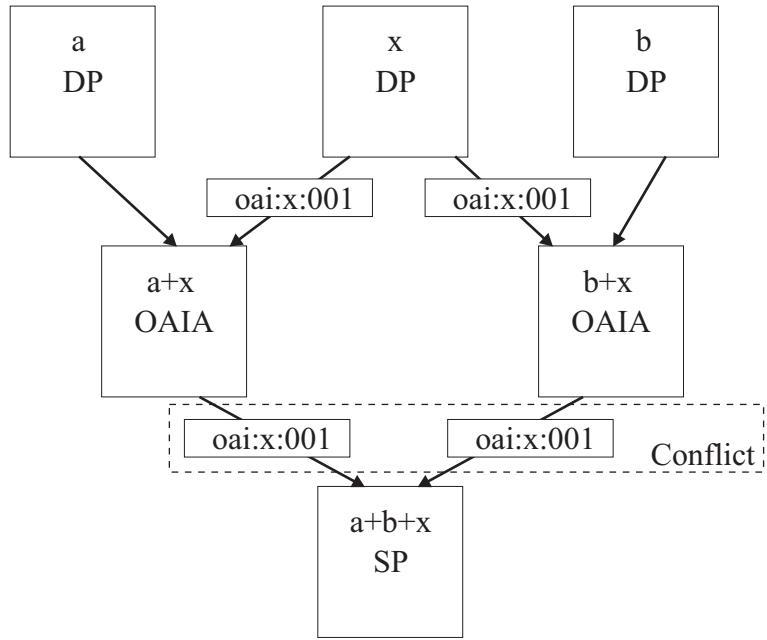}
\caption{Identifier Conflict in Hierarchical Harvesting}
\label{fig:oaia_identifier}
\end{figure}

When there is more than one path from a DP to an SP, the SP may need to resolve a collision between two or more records with the same \textit{oai-identifier}.

Figure \ref{fig:oaia_identifier} shows how one record (with a unique \textit{oaiidentifier}), may appear twice to a harvester. Three repositories, a, b, and x, are being harvested by two aggregators, a+x and b+x. When the service provider a+b+x harvests from a+x and b+x, it will get duplicates for every record from the data provider x.

To resolve collisions, a service provider can either store both records or attempt to discard one. The following  are some possible policies for record discarding:

\begin{description}
\item[Duplicate Records]If colliding records are the same, or close, the duplicates could be safely discarded.
\item[Trusted Sources]The SP in Figure \ref{fig:oaia_identifier} may, for example, trust OAIA b+x more than OAIA a+x, in which case the SP could discard or overwrite any colliding records harvested from OAIA a+x.
\item[Most Recent]It may be possible to distinguish the most recent (and hence most authoritative) record using the datestamps given by the aggregator's provenance data (e.g., OAIAs a+x and b+x in Figure \ref{fig:oaia_identifier}). 
\end{description}

\section{OAI-PMH-Gateway, Value-Added Services} \label{sect:oai_gateway}


A gateway between two resource discovery systems translates operations from one system into operations in another system. An OAI-PMH Gateway is responsible for converting OAI-PMH for use by other applications and vice-versa. Unlike the OAI-PMH Cache and Proxy, the Gateway service does not necessarily retain the original data or OAI-PMH interface.  The objective of a Gateway is to extend OAI-PMH-compliant repositories to other protocols or applications; for example:

\begin{description}
\item[Protocol Broker]A protocol broker could convert HTTP-based OAI-PMH requests to SOAP messages, or extend OAI-PMH to a Web Service model.
\item[Gateway for Crawlers] A gateway for web crawlers could translate OAI-PMH-compliant repositories to a series of linked web pages, which allows web search engines that do not support the OAI-PMH to index the ``Deep Web'' contained within OAI-PMH-compliant repositories.
\item[Value-Added Services] A gateway could cache the full-text document, and then provide value-added services, such as citation extraction, which can then be re-exposed through its own OAI-PMH interface.
\item[Subject Gateway]  A subject gateway could help build a topic-specific service by harvesting records and then exposing them by subject criteria.
\end{description}

A gateway service may create a large overhead for DPs, especially if the gateway is designed to serve machine-based applications (e.g., web crawlers). This situation is where the OAI-PMH Cache is relevant because the hierarchical structure of OAI-PMH Cache will reduce the overhead for the source DPs to a minimum. At the same time, the gateway service itself could use flow control mechanisms, such as HTTP-throttle software, to reduce the overhead.


\section{End-user Service, Search/Citation, Cross Archive Linking Service} \label{sect:end_user_service}

There are several considerations for implementing end-user services using the OAI-PMH:

\begin{description}
\item[Unique \textit{oai-identifier}]The globally unique \textit{oai-identifier} could be a basis for SP/DP interoperability.
\item[Cross-Archive Linking]With various DPs providing metadata of different qualities and formats, cross-archive linking is necessary to integrate them into one linking environment.
\item[Cross-Service Linking]Many services may be implemented on the same harvested records. A combination of these services will be very useful.
\item[Parallel Metadata Set]While OAI-PMH defines Dublin Core as the mandatory metadata format, each provider/community may implement its own metadata sets; supporting this variety of metadata sets is a challenging issue. 
\item[Online XML Support]DP supports for web presentation of proprietary XML metadata formats.
\end{description}

Different SPs can be aware of and link to each other by using OAI-PMH unique identifiers. A distributed service model could be accomplished by sharing different services.
\begin{figure}[ht]
\centering
\includegraphics[]{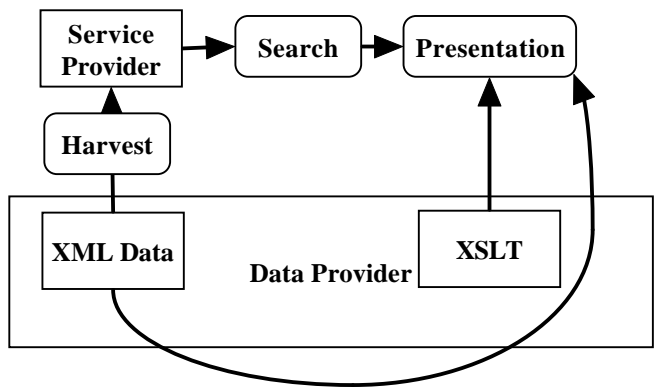}
\caption{XSLT Processing}
\label{fig:xslt}
\end{figure}

Figure \ref{fig:xslt} shows how to support parallel metadata presentation by XSLT processing. DPs will export their metadata (by OAI-PMH) and presentation format (by XSLT). An SP harvests the metadata and builds a search interface. The resource discovery is performed by the SP, and the final presentation of the data is accomplished by the DP's XSLT. With this mechanism, DPs may define an explicit method for presenting the metadata format, which is especially useful for rarely used or repository-specific metadata formats. OAI-PMH 2.0 introduces a "branding" mechanism for the SP to render the metadata in the stylesheet specified by the DP.


For research literature OAI-PMH can provide a transport mechanism for the bibliographies, as well as metadata.  The references can then be harvested by "reference-aware" services that can provide linking services (e.g.,  by generating a URL to the full-text from the bibliographic reference). Further services can be built on the linked references, such as citation analysis.  By aggregating many archives of research literature, one linking service will be able to provide reference links across many small archives.

Currently, there is no widely used XML metadata format in DPs that supports bibliographic references. Many archives, if they support references at all, will only export the text of the reference, and rely on add-on services to parse the text into bibliographic components. Other archives (e.g., Biomed Central), whose core document format is XML, will be able to provide fully marked-up references, so that only minimal processing will be required by linking services. The standardization of OpenURL, which is gathering growing support as an open and extensible system for reference linking using HTTP URLs, may result in a widely-used bibliographic-aware metadata format \cite{openurl}.

A citation-linking system is likely to be hierarchical because middle normalization layers will be needed to parse source archive reference lists into a common format that end-user services can link to bibliographic databases; furthermore references will also need to be provided in a format understood by the end-user.

\section{Experiments} \label{sect:working_applications}
\subsection{OAI-PMH-Proxy}
Our first experiment is an OAI-PMH-specific proxy that takes a URL of the format:\\
{\tt{http://foo.org/OAIProxy/\{repositoryid\}?\{oai verb\}}} \\
 Its function is to filter XML encoding errors. This proxy relies on a preexisting mapping table between an OAI-PMH repository ID and a baseURL. When an OAI-PMH request is issued, the proxy forwards the request to the corresponding data provider. The XML response is parsed by a DOM parser; if any XML encoding errors exist, the proxy tries to delete bad records based on the detailed error message from the DOM parser. The proxy then returns the corrected XML response.

\subsection{OAI Aggregation/Caching/Filtering} \label{sect:experiments_oaia}

An OAI-PMH cache service has been explored in several experiments, including OAI Aggregator, Arc, and CiteBase.  Both Arc and Citebase act as DPs  disseminating Dublin Core metadata harvested from other DPs.

OAIA is specially designed to mirror OAI repositories. OAIA creates a duplicate of all available data from the source repositories, excluding the set hierarchy (with OAI 1.1, the set hierarchy can only be ascertained through exhaustive querying of each set).

OAIA is designed to facilitate OAI-PMH Gateway services, which rely on fast, reliable access to OAI repositories.  For DP9, this consists of a few list identifiers requests, followed by large numbers of individual record requests (as a Web crawler requests the corresponding record's web page). Otherwise, this kind of behavior can lead to large numbers of (possibly unwanted) requests to source repositories exposed through DP9.

OAIA also acts as a gateway from legacy OAI-PMH implementations (1.x) to the most recent version of the OAI-PMH (2.0). As well as being able to harvest from any repository that has been OAI-PMH-compliant, OAIA converts the required Dublin Core metadata format to the most recent OAI-PMH version.

Given a new repository URL, OAIA issues an Identify request. The Identify response is stored so an SP can retrieve from OAIA the source repository's data policies, etc. A ListMetadataFormats request is issued to find out which metadata formats are supported. Each record is then requested for each metadata format (either using the batch command ListRecords, or GetRecord, depending on the repository's reliability). The metadata is stored as it was received from the source repository, ready for an SP to harvest. The record's datestamp is changed to the time the record was harvested by OAIA.

OAIA provides two views to harvesters of the records it has collected: the aggregated collection, or individual repositories by using a wrapped URL.

When an aggregated OAIA collection receives a ListMetadataFormats request, it lists all the metadata formats used by any of the harvested repositories (which may include variants of the same metadata format). When the same request is made to a wrapped repository, it lists only the metadata formats supported by that repository.


\subsection{Gateway}
DP9 is a gateway service that allows general search engines, (e.g. Google, Inktomi) to index OAI-compliant archives.  DP9 does this by providing persistent URLs for records, and converting them to OAI-PMH queries against the appropriate repository when the URL is requested. This service allows search engines that do not support the OAI-PMH to index the "Deep Web" contained within OAI-PMH-compliant repositories.

DP9 is a gateway service; it does not cache the records and only forwards requests to corresponding DPs.  This process ensures DP9 records are always up to date; however, its quality of service is highly dependent on the availability of DPs.  On the other hand, an aggressive crawler using DP9 can rapidly send requests without regard for the load it places on the DPs. The robot exclusion protocol [5] at the DP site will not be observed because the requests originate from DP9. The OAIA is used relieve the load on DPs.

Another gateway service is the reference extraction module in CiteBase. CiteBase extracts the bibliography from arXiv.org documents and exposes them by an additional OAI-PMH interface. These data are then harvested by ODU to build its own citation service.

CiteBase adds reference data by harvesting new records from a repository's OAI-PMH interface, then separately downloading the full text for parsing. The parsed bibliography is added to the existing metadata, and is used by CiteBase or harvested by other services.

The CiteBase concept could be extended from the current supported repository (arXiv.org \cite{arXiv}) to a general service for any full-text scientific repository - assuming tools can be developed to parse the bibliography.

\subsection{End-User Services}

\begin{figure}[htb]
\centering
\includegraphics[]{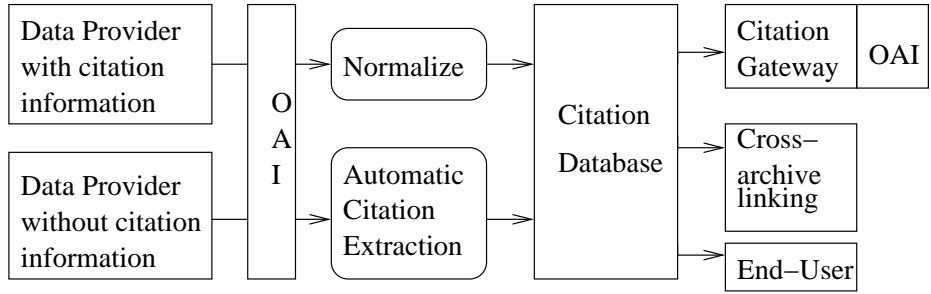}
\caption{Cross Archive Citation Link}
\label{fig:citelink}
\end{figure}

Both Archon and CiteBase implement a cross-archive search interface; Archon focuses more on harvesting heterogeneous collections and builds an interactive search interface based on harvested metadata, and CiteBase concentrates on automatic reference extraction. Both applications may harvest from the same repository (e.g. arXiv.org \cite{arXiv}) and implement different services for the same record. With the quick adoption of OAI-PMH, we believe this will become a common situation. We implemented two prototypes for cross-linking between Archon and CiteBase (Figure \ref{fig:citelink}).

The first approach is to reexpose value-added metadata through an OAI-PMH interface. By this method Archon harvests citation data from CiteBase, APS, CERN and other sources, it then builds a cross archive linking service such as a citation in APS may lead to document in CiteBase and vice-versa. Another prototype is based on dynamic linking: both services link to a broker page, and the broker page dynamically checks whether a service exists for a specific record. If so, it adds a link to the corresponding SP. In order to know which records are available in advance, the broker issues an OAI GetRecord lookup to the target service (which has an OAI-PMH export). Based on the reply, the broker knows whether a record is harvested. We envision that a DP may also link to this broker page for additional services for its data.

\section{Summary} \label{sect:summary}
The OAI-PMH is an important infrastructure for supporting distributed networked information services. This paper introduces the concepts of OAI-PMH proxies, OAI-PMH caches, OAI-PMH gateways as tools for the optimization of the functioning of the DP/SP model underlying the OAI-PMH. To demonstrate the usability of this framework, we have built several prototype services. These demonstration systems and source codes are available at the web sites of both ODU and the Southampton group. 
\section{Acknowledgements}
Our thanks to Herbert Van de Sompel and Steve Hitchcock for providing valuable input and reviewing the manuscript.
\bibliography{oaiarch}
\bibliographystyle{plain}

\end{document}